\documentclass[11pt,a4paper]{article}
\usepackage[T1]{fontenc}
\usepackage[utf8]{inputenc}
\usepackage{graphicx}
\usepackage{amsmath,amssymb}
\usepackage{cite}
\usepackage[colorlinks=true,linkcolor=blue,citecolor=blue,urlcolor=blue]{hyperref}
\usepackage{orcidlink}
\usepackage{float}
\usepackage{QCDEX2026}
\newcommand{\PaperTitle}{$Z$-Boson Production in Proton--Lead Collisions with MRW Nuclear TMDs}

\newcommand{\PaperAbstract}{%
The production of dileptons in proton--lead collisions probes the transverse-momentum structure of partons in a nuclear environment. We summarize a $k_t$-factorization study of $Z$ boson production at $\sqrt{s_{\mathrm{NN}}}=8.16\,\mathrm{TeV}$, comparing transverse-momentum-dependent (TMD) distributions constructed with leading-order and Next-to-leading-order Martin--Ryskin--Watt (LO-MRW and NLO-MRW) prescriptions. Parton-level predictions obtained with \textsc{KaTie} are confronted with the CMS measurements in two invariant-mass regions, using the dilepton transverse momentum, $\phi^*_\eta$ and center-of-mass rapidity. The models describe the overall trends near the $Z$ resonance, while the virtuality constraint in NLO-MRW changes the high-transverse-momentum behavior. At lower dilepton masses, both prescriptions underestimate parts of the data, particularly the NLO-MRW result. These comparisons illustrate both the potential and the limitations of last-step-evolution nuclear TMD models for $Z$ boson production in proton--nucleus collisions.%
}

\begin{document}
\noindent\includegraphics[width=0.4\textwidth]{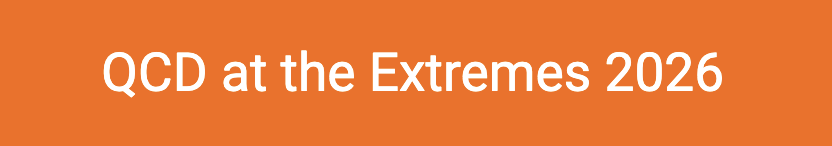}\par
\vspace{0.6em}
\begin{center}
{\LARGE\bfseries \PaperTitle\par}
\vspace{0.8em}
{\large S.~Rezaie$^{1,*}$\,\orcidlink{0000-0002-1852-1619},\quad
and K.~Azizi$^{1,2,3}$\,\orcidlink{0000-0003-3741-2167}\par}
\vspace{0.5em}
{\small
$^1$School of Particles and Accelerators, Institute for Research in Fundamental Sciences (IPM), Tehran, Iran\par
$^2$Department of Physics, University of Tehran, Tehran, Iran\par
$^3$Department of Physics, Dogus University, Istanbul, Turkey\par}
\vspace{0.35em}
{\small $^*$Speaker and corresponding author: \href{mailto:somayeh@ipm.ir}{somayeh@ipm.ir}\par}
\vspace{0.35em}
{\small\itshape Contribution to QCD at the Extremes 2026\par}
\end{center}
\vspace{0.5em}
\begin{abstract}
\PaperAbstract
\end{abstract}
\section{Introduction}

$Z$ boson production provides a clean probe of partonic dynamics in hadronic and nuclear collisions. Because a $Z$ boson does not interact strongly with the produced medium, its leptonic decay products retain information about the initial state. In proton--lead (p--Pb) collisions, measurements of the Drell--Yan process can therefore constrain the interplay between nuclear parton densities, partonic transverse momentum, and the hard scattering. The CMS Collaboration has measured differential dimuon cross sections at $\sqrt{s_{\mathrm{NN}}}=8.16\,\mathrm{TeV}$, including the region around the $Z$ resonance and a lower invariant-mass interval \cite{CMS:2021}.

In collinear factorization, incoming partons carry longitudinal momentum fractions but no resolved transverse momentum at the hard-scattering level. The $k_t$-factorization approach instead retains their transverse momentum and combines transverse-momentum-dependent (TMD) parton distributions with off-shell partonic matrix elements. This is especially relevant for the dilepton transverse-momentum spectrum and related angular observables \cite{vanHameren:2018,Blanco:2019}. A central modeling issue is how to construct reliable nuclear TMDs from more familiar collinear PDFs.

This contribution summarizes our study \cite{Rezaie:2025} presented at \emph{QCD at the Extremes 2026}. We compare two Martin--Ryskin--Watt (MRW) prescriptions, denoted LO-MRW and NLO-MRW, and contrast their predictions with CMS data and an existing parton-branching (PB) calculation. The emphasis is on what the comparison teaches us about angular versus virtuality ordering, the relative role of partonic channels, and the low-mass region where missing contributions become important. The detailed derivations and a broader collection of results are given in Ref.~\cite{Rezaie:2025}.

\section{\texorpdfstring{$k_t$}{kt}-factorization and MRW nuclear TMDs}

Schematically, the factorized cross section for a hard final state $X$ takes the form
\begin{equation}
 \sigma_{p\mathrm{Pb}\to X}
 =\sum_{a,b}\int\frac{dx_1}{x_1}\frac{dx_2}{x_2}
 \frac{dk_{1t}^{2}}{k_{1t}^{2}}\frac{dk_{2t}^{2}}{k_{2t}^{2}}
 {f}_{a}(x_1,k_{1t}^{2},\mu^{2})
 {f}_{b}(x_2,k_{2t}^{2},\mu^{2})
 \hat\sigma^{*}_{ab\to X} .
 \label{eq:ktfac}
\end{equation}
Here $\hat\sigma^*$ denotes the off-shell hard-partonic cross section, and the TMDs are defined consistently with the integration measure in Eq.~\eqref{eq:ktfac}. The proton and lead inputs are evolved collinear parton densities supplemented by prescriptions that generate transverse momentum. In the numerical study, the proton PDF input is CT18NLO and the lead input is the nCTEQ15 nuclear set \cite{CT18,nCTEQ15,Rezaie:2025}.

\subsection{LO-MRW: the last real emission}

In the LO-MRW construction, the transverse momentum of the incoming parton is generated at the last resolvable branching. A generic quark or gluon TMD can be represented schematically as \cite{MRW:2010}
\begin{equation}
 {f}_a(x,k_t^2,\mu^2)
 =T_a(k_t^2,\mu^2)\,
 \frac{\alpha_s(k_t^2)}{2\pi}
 \sum_b\int_x^1 dz\,
 P_{ab}(z)\,f_b\!\left(\frac{x}{z},k_t^2\right)
 \Theta^{\delta^{{ab}}}(z_{max}-z),
 \label{eq:lomrw}
\end{equation}
where $f_b$ is the momentum-weighted collinear density in the convention of Ref.~\cite{Rezaie:2025}. The Sudakov factor $T_a$ gives the no-resolvable-emission probability between the last-branching scale and the factorization scale. The angular-ordering (AO) restriction regulates the soft-gluon region. In the formulation used here, the characteristic cutoff is $z_{\max}=\mu/(\mu+k_t)$ for the relevant diagonal contributions. Angular ordering permits a broader transverse-momentum tail than imposing $k_t<\mu$ everywhere.

\subsection{NLO-MRW: a virtuality-ordered scale choice}

The NLO-MRW prescription introduces the virtuality of the last branching as
\begin{equation}
 k^2=\frac{k_t^2}{1-z},\qquad
 \Theta(\mu^2-k^2),
 \label{eq:virt}
\end{equation}
with $\alpha_s$, the collinear density, and the Sudakov factor evaluated at the virtuality scale. The additional restriction $k^2\le\mu^2$ reduces the allowed $z$ interval, particularly at large $k_t$. For the practical implementation studied in Ref.~\cite{Rezaie:2025}, this virtuality-ordered prescription is evaluated using LO real splitting kernels together with higher-order PDF inputs.

The PB model used as a comparison is not a third MRW variant: it builds the transverse-momentum distribution through an iterative branching evolution. We use its published results as a reference rather than recomputing them \cite{Blanco:2019,Rezaie:2025}.

\section{Numerical setup and observables}

Parton-level events are calculated with \textsc{KaTie}, which provides matrix elements for incoming off-shell partons \cite{vanHameren:2018}. We consider the $Z$ boson contribution to $p\mathrm{Pb}\to\mu^+\mu^-+X$ and include the relevant quark--antiquark and quark--gluon channels, schematically $q\bar q\to\ell^+\ell^-$, $q\bar q\to\ell^+\ell^-g$, and $qg\to\ell^+\ell^-q$ (together with charge-conjugate channels). Five light and heavy active quark flavors are included, as in the original study. The common central renormalization and factorization scales are set to
\begin{equation}
 \mu_R=\mu_F=\sqrt{m_{\ell\ell}^{2}+\bigl(p_T^{\ell\ell}\bigr)^2}.
 \label{eq:scale}
\end{equation}
The generated predictions are evaluated in the nucleon--nucleon center-of-mass frame and compared with CMS fiducial measurements for $|\eta_{\mu}^{\mathrm{lab}}|<2.4$, with the transverse-momentum selection taken from the corresponding experimental mass-region definition \cite{CMS:2021,Rezaie:2025}. The analyzed integrated luminosity is $173\,\mathrm{nb}^{-1}$.

We study the dilepton transverse momentum $p_T^{\ell\ell}$, the dilepton center-of-mass rapidity $y_{\mathrm{CM}}$, and the angular variable $\phi^*_\eta$. The last observable is approximately related to the ratio $p_T^{\ell\ell}/m_{\ell\ell}$ at small transverse momentum and is defined by
\begin{equation}
 \phi^*_\eta=\tan\!\left(\frac{\pi-\Delta\phi}{2}\right)
 \sin\theta^*_\eta,
  \label{eq:phistar}
\end{equation}
where $\Delta\phi$ is the azimuthal separation of the leptons. We discuss two dimuon mass intervals: $60<m_{\mu\mu}<120\,\mathrm{GeV}$, which contains the $Z$ resonance, and $15<m_{\mu\mu}<60\,\mathrm{GeV}$, where off-resonance Drell--Yan production is more important.

\section{Results and discussion}

\subsection{Upper dimuon masses}

\begin{figure}[!htbp]
 \centering
 \includegraphics[width=0.33\linewidth]{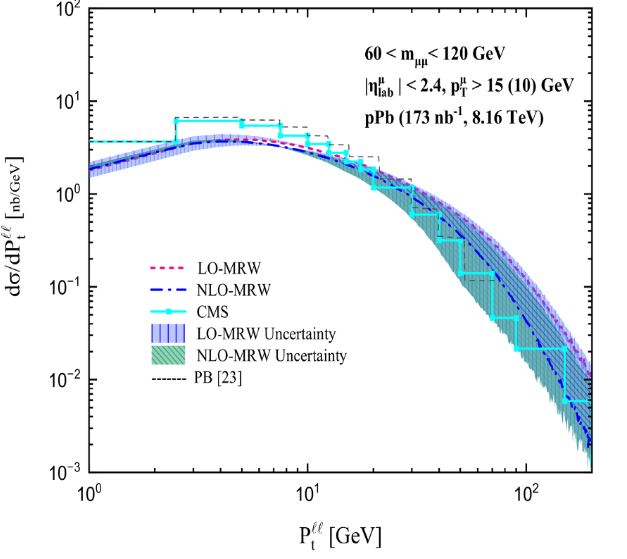}
 \includegraphics[width=0.35\linewidth]{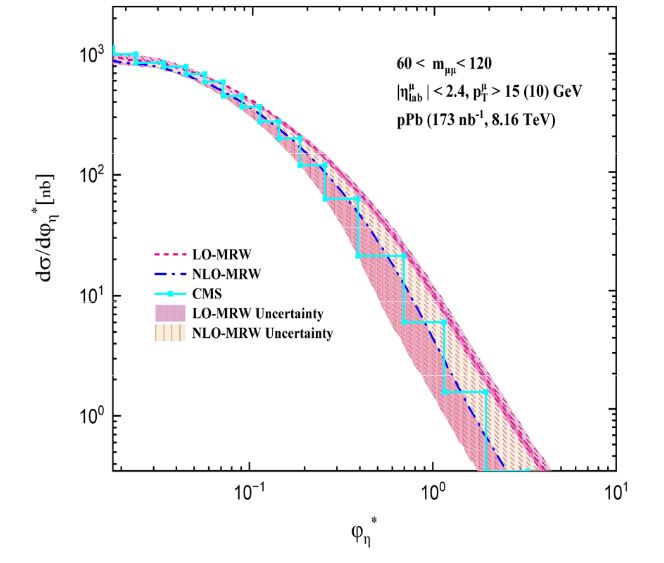}
 \includegraphics[width=0.35\linewidth]{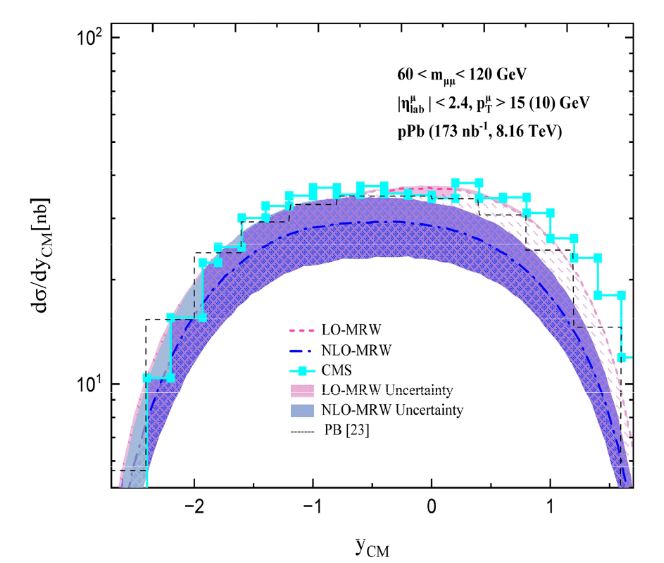}
 \caption{Differential dimuon cross sections in p--Pb collisions at $\sqrt{s_{\mathrm{NN}}}=8.16\,\mathrm{TeV}$ for $60<m_{\mu\mu}<120\,\mathrm{GeV}$. The upper-left, upper-right, and lower panels show $p_T^{\ell\ell}$, $\phi^*_\eta$ and $y_{\mathrm{CM}}$, respectively. LO-MRW and NLO-MRW results are compared with CMS data and the PB reference. Panels reassembled from Fig.~3 of Ref.~\cite{Rezaie:2025}.}
 \label{fig:high}
\end{figure}

Figure~\ref{fig:high} compares LO-MRW, NLO-MRW, and the published PB baseline with the CMS data in the $60<m_{\mu\mu}<120\,\mathrm{GeV}$ interval. All three approaches describe the broad shape of the $p_T^{\ell\ell}$ spectrum and of $\phi^*_\eta$. The transverse-momentum distributions are particularly informative because the difference between the two MRW prescriptions grows in regions sensitive to the final branching. The comparatively strong virtuality constraint in NLO-MRW suppresses large transverse momenta, whereas LO-MRW retains a larger tail. Consequently, the predictions can separate even when they give similar results near the resonance \cite{Rezaie:2025}.

The rapidity distribution is generally described more consistently than the tails of the transverse observables. At the largest $p_T^{\ell\ell}$ or $\phi^*_\eta$ values, some LO-MRW predictions overshoot the data, while NLO-MRW can lie closer to the measurements. These trends should not be read as a uniform preference for a single prescription: agreement varies across bins, the plotted bands reflect scale sensitivity, and the result also depends on which hard subprocesses are included. The three observables thus test complementary parts of the nuclear TMD model.

The channel decomposition given in Ref.~\cite{Rezaie:2025} shows that the quark--antiquark-induced contributions play a leading role in the selected distributions. Additional channels involving a gluon nevertheless affect their magnitude and shape, especially away from the lowest transverse momenta. The contrast between the MRW prescriptions is therefore not only a matter of choosing a collinear nuclear PDF: their different final-branching constraints are directly visible in the computed observables.

\subsection{Lower dimuon masses}

\begin{figure}[H]
 \centering
 \includegraphics[width=0.33\linewidth]{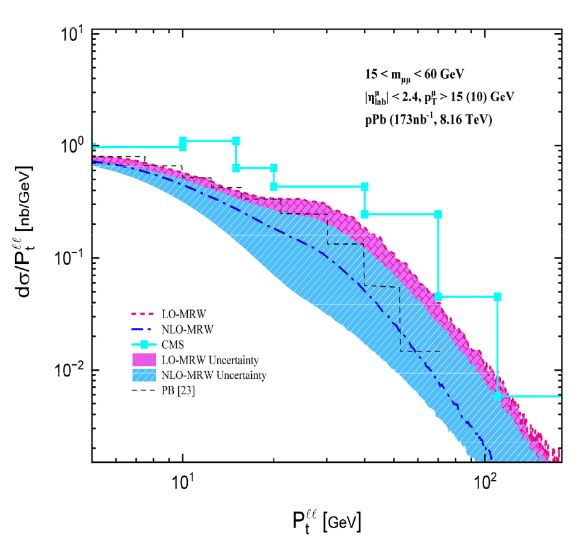}
 \includegraphics[width=0.35\linewidth]{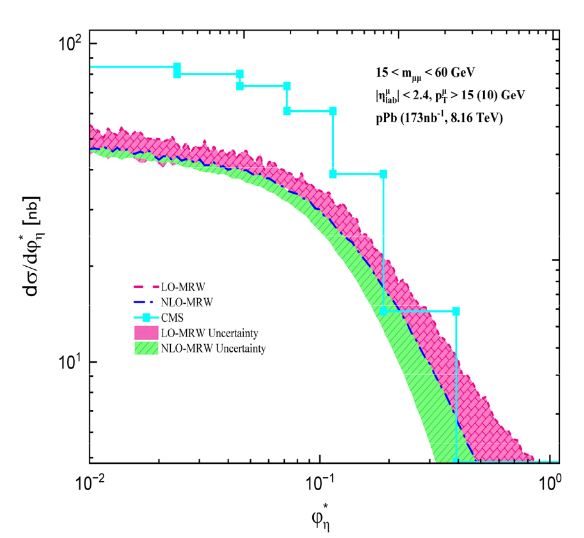}
 \includegraphics[width=0.35\linewidth]{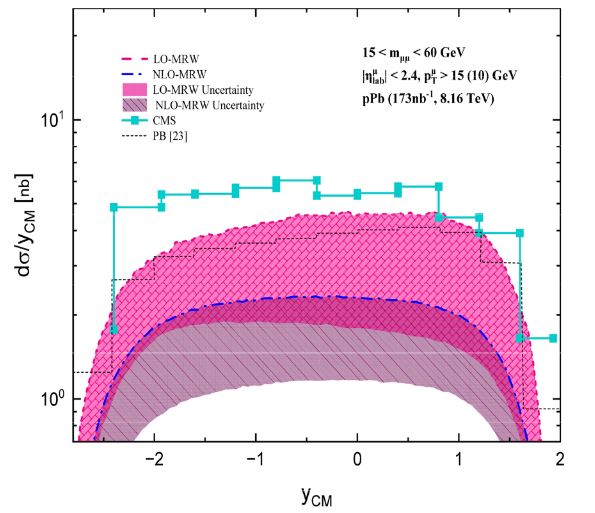}
 \caption{As in Fig.~\ref{fig:high}, but for the lower-mass interval $15<m_{\mu\mu}<60\,\mathrm{GeV}$. The three panels display $p_T^{\ell\ell}$, $\phi^*_\eta$ and $y_{\mathrm{CM}}$. Panels reassembled from Fig.~5 of Ref.~\cite{Rezaie:2025}.}
 \label{fig:low}
\end{figure}

The lower-mass comparison in Fig.~\ref{fig:low} exposes a more pronounced shortfall in the model predictions. In the interval $15<m_{\mu\mu}<60\,\mathrm{GeV}$, both MRW calculations tend to underestimate the CMS cross sections, with the NLO-MRW curves generally lower than the LO-MRW ones. This behavior is visible in the $p_T^{\ell\ell}$ and $y_{\mathrm{CM}}$ distributions and is particularly striking when the central predictions are compared with the measurements. The PB reference also lies below the data in several parts of this interval \cite{Rezaie:2025}.

This discrepancy is not, by itself, evidence that the nuclear TMD evolution is incorrect. The calculation is based on a restricted set of leading hard subprocesses and does not provide a complete higher-order description of low-mass Drell--Yan production. Additional hard radiation, corrections to the off-shell matrix elements, and a more complete treatment of perturbative matching may alter the normalization. The data consequently motivate further theoretical improvements rather than a direct rejection of one or the other TMD prescription.

\section{Conclusions}

We have presented a compact overview of $Z$ boson production in p--Pb collisions within $k_t$-factorization using LO-MRW and NLO-MRW nuclear TMDs. Near the $Z$ resonance, the calculations reproduce the overall trends of the CMS dilepton data, with the strongest model differences appearing in transverse-momentum-sensitive regions. The virtuality restriction of NLO-MRW produces a more suppressed high-$k_t$ contribution than LO-MRW, underscoring the significance of the ordering prescription used to construct the TMDs.

At lower invariant masses, the two MRW schemes substantially underestimate portions of the measured cross sections; the truncated set of hard subprocesses and the absence of a complete higher-order matching are important caveats. Further work should examine additional channels, controlled perturbative corrections, and applications to lead--lead collisions. A complete presentation of the formalism and numerical analysis is available in Ref.~\cite{Rezaie:2025}.

\section*{Acknowledgements}
We thank the organizers of \emph{QCD at the Extremes 2026} for the opportunity to present and discuss this work. We acknowledge support from the Institute for Research in Fundamental Sciences (IPM).

\begingroup
\footnotesize
\bibliographystyle{QCDEX2026}
\bibliography{QCDEX2026}
\endgroup
\end{document}